 \documentclass[preprint,prd,tightenlines,superscriptaddress]{revtex4-1}

\usepackage{graphicx} 
\usepackage{dcolumn}  
\usepackage{colordvi}
\usepackage{color}
\usepackage{epstopdf}
\usepackage{pstricks}
\usepackage{amssymb}
\usepackage{url}
\graphicspath{{ps}}
\usepackage{hyperref}
\usepackage{tabularx}
\usepackage{multirow}
\usepackage{units}
\usepackage{siunitx}
\usepackage{hyphenat}
\usepackage{subfigure}
\usepackage[italic]{hepnames}

\renewcommand{\PBzero}{\ensuremath{\HepParticle{\PB}{}{}^0}\xspace}

\renewcommand{\Pgpz}{\ensuremath{\HepParticle{\Pgp}{}{}^0}\xspace}

\renewcommand{\PKzS}{\ensuremath{\HepParticle{\PK}{}{}^0_{\rm S}}\xspace}

\input ./belle2sym.tex

\begin{document}

\def\belletwo {\it {Belle II}}

\vspace*{-3\baselineskip}
\resizebox{!}{3cm}{\includegraphics{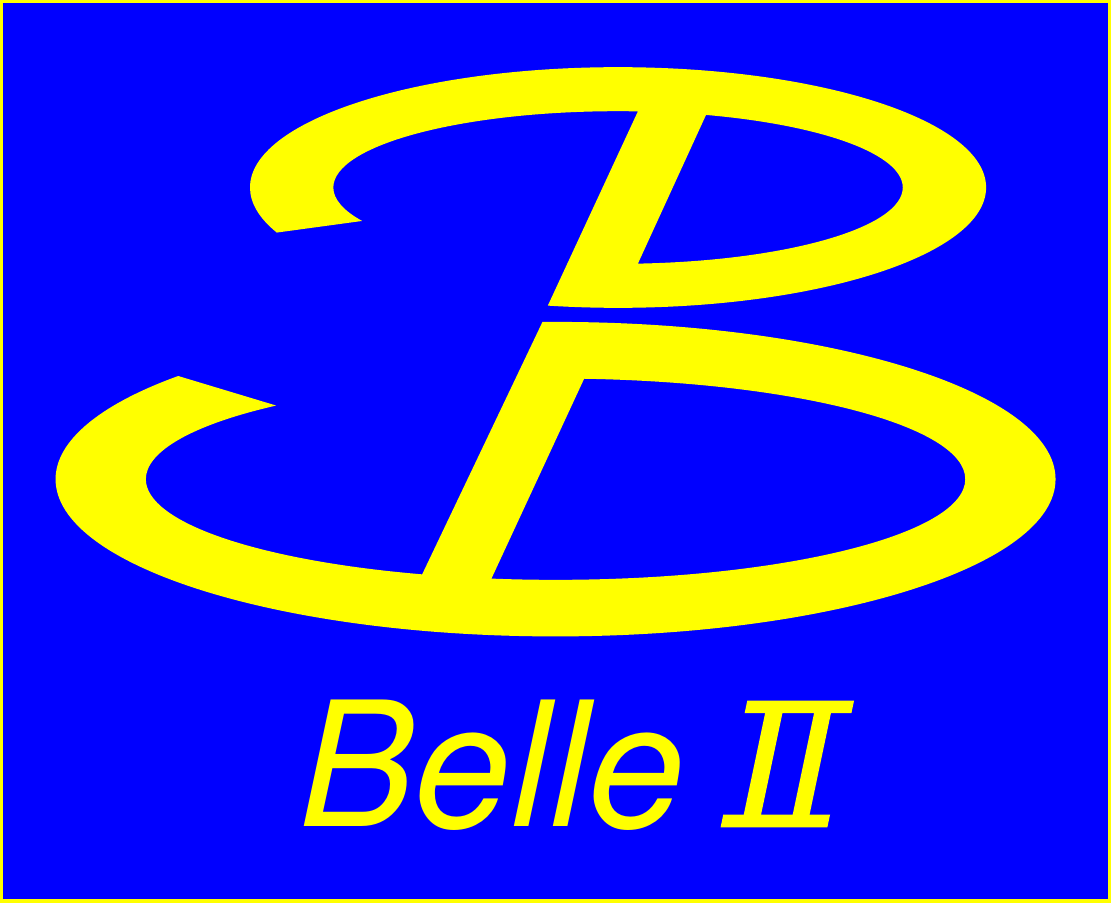}}

\vspace*{-5\baselineskip}
\begin{flushright}
BELLE2-CONF-PH-2020-001
\\
\today
\end{flushright}

\title { \quad\\[0.5cm] Charmless $B$ decay reconstruction in 2019 Belle II data}

\newcommand{\instSinica}{Academia Sinica, Taipei 11529, Taiwan}
\newcommand{\instBeihang}{Beihang University, Beijing 100191, China}
\newcommand{\instBUAP}{Benemerita Universidad Autonoma de Puebla, Puebla 72570, Mexico}
\newcommand{\instBNL}{Brookhaven National Laboratory, Upton, New York 11973, U.S.A.}
\newcommand{\instBINP}{Budker Institute of Nuclear Physics SB RAS, Novosibirsk 630090, Russian Federation}
\newcommand{\instCMU}{Carnegie Mellon University, Pittsburgh, Pennsylvania 15213, U.S.A.}
\newcommand{\instCPPM}{Centre de Physique des Particules de Marseille, 13288 Marseille, France}
\newcommand{\instCinvestavIPN}{Centro de Investigacion y de Estudios Avanzados del Instituto Politecnico Nacional, Mexico City 07360, Mexico}
\newcommand{\instPrague}{Faculty of Mathematics and Physics, Charles University, 121 16 Prague, Czech Republic}
\newcommand{\instChiangMai}{Chiang Mai University, Chiang Mai 50202, Thailand}
\newcommand{\instChiba}{Chiba University, Chiba 263-8522, Japan}
\newcommand{\instChonnam}{Chonnam National University, Gwangju 61186, South Korea}
\newcommand{\instConacyt}{Consejo Nacional de Ciencia y Tecnolog\'{\i}a, Mexico City 03940, Mexico}
\newcommand{\instDESY}{Deutsches Elektronen--Synchrotron, 22607 Hamburg, Germany}
\newcommand{\instDuke}{Duke University, Durham, North Carolina 27708, U.S.A.}
\newcommand{\instDuyTan}{Institute of Theoretical and Applied Research (ITAR), Duy Tan University, Hanoi 100000, Vietnam}
\newcommand{\instENEA}{ENEA Casaccia, I-00123 Roma, Italy}
\newcommand{\instEri}{Earthquake Research Institute, University of Tokyo, Tokyo 113-0032, Japan}
\newcommand{\instJuelich}{Forschungszentrum J\"{u}lich, 52425 J\"{u}lich, Germany}
\newcommand{\instFuJen}{Department of Physics, Fu Jen Catholic University, Taipei 24205, Taiwan}
\newcommand{\instFudan}{Key Laboratory of Nuclear Physics and Ion-beam Application (MOE) and Institute of Modern Physics, Fudan University, Shanghai 200443, China}
\newcommand{\instGoettingen}{II. Physikalisches Institut, Georg-August-Universit\"{a}t G\"{o}ttingen, 37073 G\"{o}ttingen, Germany}
\newcommand{\instGifu}{Gifu University, Gifu 501-1193, Japan}
\newcommand{\instSOKENDAI}{The Graduate University for Advanced Studies (SOKENDAI), Hayama 240-0193, Japan}
\newcommand{\instGyeongsang}{Gyeongsang National University, Jinju 52828, South Korea}
\newcommand{\instHanyang}{Department of Physics and Institute of Natural Sciences, Hanyang University, Seoul 04763, South Korea}
\newcommand{\instKEK}{High Energy Accelerator Research Organization (KEK), Tsukuba 305-0801, Japan}
\newcommand{\instJPARC}{J-PARC Branch, KEK Theory Center, High Energy Accelerator Research Organization (KEK), Tsukuba 305-0801, Japan}
\newcommand{\instHSE}{Higher School of Economics (HSE), Moscow 101000, Russian Federation}
\newcommand{\instIISER}{Indian Institute of Science Education and Research Mohali, SAS Nagar, 140306, India}
\newcommand{\instIITBhubaneswar}{Indian Institute of Technology Bhubaneswar, Satya Nagar 751007, India}
\newcommand{\instIITGuwahati}{Indian Institute of Technology Guwahati, Assam 781039, India}
\newcommand{\instIITHyderabad}{Indian Institute of Technology Hyderabad, Telangana 502285, India}
\newcommand{\instIITMadras}{Indian Institute of Technology Madras, Chennai 600036, India}
\newcommand{\instIndiana}{Indiana University, Bloomington, Indiana 47408, U.S.A.}
\newcommand{\instIHEPRussia}{Institute for High Energy Physics, Protvino 142281, Russian Federation}
\newcommand{\instHEPHYVienna}{Institute of High Energy Physics, Vienna 1050, Austria}
\newcommand{\instIHEPChina}{Institute of High Energy Physics, Chinese Academy of Sciences, Beijing 100049, China}
\newcommand{\instChennai}{Institute of Mathematical Sciences, Chennai 600113, India}
\newcommand{\instIPP}{Institute of Particle Physics (Canada), Victoria, British Columbia V8W 2Y2, Canada}
\newcommand{\instIOP}{Institute of Physics, Vietnam Academy of Science and Technology (VAST), Hanoi, Vietnam}
\newcommand{\instIFIC}{Instituto de Fisica Corpuscular, Paterna 46980, Spain}
\newcommand{\instFrascati}{INFN Laboratori Nazionali di Frascati, I-00044 Frascati, Italy}
\newcommand{\instNapoliINFN}{INFN Sezione di Napoli, I-80126 Napoli, Italy}
\newcommand{\instPadovaINFN}{INFN Sezione di Padova, I-35131 Padova, Italy}
\newcommand{\instPerugiaINFN}{INFN Sezione di Perugia, I-06123 Perugia, Italy}
\newcommand{\instPisaINFN}{INFN Sezione di Pisa, I-56127 Pisa, Italy}
\newcommand{\instRomaINFN}{INFN Sezione di Roma, I-00185 Roma, Italy}
\newcommand{\instRomaTreINFN}{INFN Sezione di Roma Tre, I-00146 Roma, Italy}
\newcommand{\instTorinoINFN}{INFN Sezione di Torino, I-10125 Torino, Italy}
\newcommand{\instTriesteINFN}{INFN Sezione di Trieste, I-34127 Trieste, Italy}
\newcommand{\instJAEA}{Advanced Science Research Center, Japan Atomic Energy Agency, Naka 319-1195, Japan}
\newcommand{\instMainz}{Johannes Gutenberg-Universit\"{a}t Mainz, Institut f\"{u}r Kernphysik, D-55099 Mainz, Germany}
\newcommand{\instGiessen}{Justus-Liebig-Universit\"{a}t Gie\ss{}en, 35392 Gie\ss{}en, Germany}
\newcommand{\instKarlsruhe}{Institut f\"{u}r Experimentelle Teilchenphysik, Karlsruher Institut f\"{u}r Technologie, 76131 Karlsruhe, Germany}
\newcommand{\instKennesaw}{Kennesaw State University, Kennesaw, Georgia 30144, U.S.A.}
\newcommand{\instKitasato}{Kitasato University, Sagamihara 252-0373, Japan}
\newcommand{\instKISTI}{Korea Institute of Science and Technology Information, Daejeon 34141, South Korea}
\newcommand{\instKorea}{Korea University, Seoul 02841, South Korea}
\newcommand{\instKSU}{Kyoto Sangyo University, Kyoto 603-8555, Japan}
\newcommand{\instKyotoU}{Kyoto University, Kyoto 606-8501, Japan}
\newcommand{\instKyungpook}{Kyungpook National University, Daegu 41566, South Korea}
\newcommand{\instLAL}{Universit\'{e} Paris-Saclay, CNRS/IN2P3, IJCLab, 91405 Orsay, France}
\newcommand{\instLPI}{P.N. Lebedev Physical Institute of the Russian Academy of Sciences, Moscow 119991, Russian Federation}
\newcommand{\instLNNU}{Liaoning Normal University, Dalian 116029, China}
\newcommand{\instLMU}{Ludwig Maximilians University, 80539 Munich, Germany}
\newcommand{\instLuther}{Luther College, Decorah, Iowa 52101, U.S.A.}
\newcommand{\instMNITJaipur}{Malaviya National Institute of Technology Jaipur, Jaipur 302017, India}
\newcommand{\instMPP}{Max-Planck-Institut f\"{u}r Physik, 80805 M\"{u}nchen, Germany}
\newcommand{\instMPGHLL}{Semiconductor Laboratory of the Max Planck Society, 81739 M\"{u}nchen, Germany}
\newcommand{\instMcGill}{McGill University, Montr\'{e}al, Qu\'{e}bec, H3A 2T8, Canada}
\newcommand{\instMETU}{Middle East Technical University, 06531 Ankara, Turkey}
\newcommand{\instMEPhI}{Moscow Physical Engineering Institute, Moscow 115409, Russian Federation}
\newcommand{\instNagoya}{Graduate School of Science, Nagoya University, Nagoya 464-8602, Japan}
\newcommand{\instNagoyaKMI}{Kobayashi-Maskawa Institute, Nagoya University, Nagoya 464-8602, Japan}
\newcommand{\instNaraWu}{Nara Women's University, Nara 630-8506, Japan}
\newcommand{\instUNAM}{National Autonomous University of Mexico, Mexico City, Mexico}
\newcommand{\instNTUTaiwan}{Department of Physics, National Taiwan University, Taipei 10617, Taiwan}
\newcommand{\instNUUTaiwan}{National United University, Miao Li 36003, Taiwan}
\newcommand{\instKrakow}{H. Niewodniczanski Institute of Nuclear Physics, Krakow 31-342, Poland}
\newcommand{\instNiigata}{Niigata University, Niigata 950-2181, Japan}
\newcommand{\instNSU}{Novosibirsk State University, Novosibirsk 630090, Russian Federation}
\newcommand{\instOkinawa}{Okinawa Institute of Science and Technology, Okinawa 904-0495, Japan}
\newcommand{\instOsakaCity}{Osaka City University, Osaka 558-8585, Japan}
\newcommand{\instRCNP}{Research Center for Nuclear Physics, Osaka University, Osaka 567-0047, Japan}
\newcommand{\instPNNL}{Pacific Northwest National Laboratory, Richland, Washington 99352, U.S.A.}
\newcommand{\instPanjab}{Panjab University, Chandigarh 160014, India}
\newcommand{\instPeking}{Peking University, Beijing 100871, China}
\newcommand{\instPanjabPAU}{Punjab Agricultural University, Ludhiana 141004, India}
\newcommand{\instRIKEN}{Theoretical Research Division, Nishina Center, RIKEN, Saitama 351-0198, Japan}
\newcommand{\instXavier}{St. Francis Xavier University, Antigonish, Nova Scotia, B2G 2W5, Canada}
\newcommand{\instSeoul}{Seoul National University, Seoul 08826, South Korea}
\newcommand{\instShandong}{Shandong University, Jinan 250100, China}
\newcommand{\instSPU}{Showa Pharmaceutical University, Tokyo 194-8543, Japan}
\newcommand{\instSoochow}{Soochow University, Suzhou 215006, China}
\newcommand{\instSoongsil}{Soongsil University, Seoul 06978, South Korea}
\newcommand{\instLjubljanaJSI}{J. Stefan Institute, 1000 Ljubljana, Slovenia}
\newcommand{\instKyiv}{Taras Shevchenko National Univ. of Kiev, Kiev, Ukraine}
\newcommand{\instTata}{Tata Institute of Fundamental Research, Mumbai 400005, India}
\newcommand{\instTUM}{Department of Physics, Technische Universit\"{a}t M\"{u}nchen, 85748 Garching, Germany}
\newcommand{\instECUTUM}{Excellence Cluster Universe, Technische Universit\"{a}t M\"{u}nchen, 85748 Garching, Germany}
\newcommand{\instTelAviv}{Tel Aviv University, School of Physics and Astronomy, Tel Aviv, 69978, Israel}
\newcommand{\instToho}{Toho University, Funabashi 274-8510, Japan}
\newcommand{\instTohoku}{Department of Physics, Tohoku University, Sendai 980-8578, Japan}
\newcommand{\instTitech}{Tokyo Institute of Technology, Tokyo 152-8550, Japan}
\newcommand{\instTokyoMetropolitan}{Tokyo Metropolitan University, Tokyo 192-0397, Japan}
\newcommand{\instUAS}{Universidad Autonoma de Sinaloa, Sinaloa 80000, Mexico}
\newcommand{\instNapoliUNIV}{Dipartimento di Scienze Fisiche, Universit\`{a} di Napoli Federico II, I-80126 Napoli, Italy}
\newcommand{\instNapoliUNIVA}{Dipartimento di Agraria, Universit\`{a} di Napoli Federico II, I-80055 Portici (NA), Italy}
\newcommand{\instPadovaUNIV}{Dipartimento di Fisica e Astronomia, Universit\`{a} di Padova, I-35131 Padova, Italy}
\newcommand{\instPerugiaUNIV}{Dipartimento di Fisica, Universit\`{a} di Perugia, I-06123 Perugia, Italy}
\newcommand{\instPisaUNIV}{Dipartimento di Fisica, Universit\`{a} di Pisa, I-56127 Pisa, Italy}
\newcommand{\instRomaUNIV}{Universit\`{a} di Roma ``La Sapienza,'' I-00185 Roma, Italy}
\newcommand{\instRomaTreUNIV}{Dipartimento di Matematica e Fisica, Universit\`{a} di Roma Tre, I-00146 Roma, Italy}
\newcommand{\instTorinoUNIV}{Dipartimento di Fisica, Universit\`{a} di Torino, I-10125 Torino, Italy}
\newcommand{\instTriesteUNIV}{Dipartimento di Fisica, Universit\`{a} di Trieste, I-34127 Trieste, Italy}
\newcommand{\instMontreal}{Universit\'{e} de Montr\'{e}al, Physique des Particules, Montr\'{e}al, Qu\'{e}bec, H3C 3J7, Canada}
\newcommand{\instIPHC}{Universit\'{e} de Strasbourg, CNRS, IPHC, UMR 7178, 67037 Strasbourg, France}
\newcommand{\instAdelaide}{Department of Physics, University of Adelaide, Adelaide, South Australia 5005, Australia}
\newcommand{\instBonn}{University of Bonn, 53115 Bonn, Germany}
\newcommand{\instUBC}{University of British Columbia, Vancouver, British Columbia, V6T 1Z1, Canada}
\newcommand{\instCincinnati}{University of Cincinnati, Cincinnati, Ohio 45221, U.S.A.}
\newcommand{\instFlorida}{University of Florida, Gainesville, Florida 32611, U.S.A.}
\newcommand{\instHamburg}{University of Hamburg, 20148 Hamburg, Germany}
\newcommand{\instHawaii}{University of Hawaii, Honolulu, Hawaii 96822, U.S.A.}
\newcommand{\instHeidelberg}{University of Heidelberg, 68131 Mannheim, Germany}
\newcommand{\instLjubljanaUniLJ}{Faculty of Mathematics and Physics, University of Ljubljana, 1000 Ljubljana, Slovenia}
\newcommand{\instLouisville}{University of Louisville, Louisville, Kentucky 40292, U.S.A.}
\newcommand{\instMalaya}{National Centre for Particle Physics, University Malaya, 50603 Kuala Lumpur, Malaysia}
\newcommand{\instLjubljanaUM}{University of Maribor, 2000 Maribor, Slovenia}
\newcommand{\instMelbourne}{School of Physics, University of Melbourne, Victoria 3010, Australia}
\newcommand{\instMississippi}{University of Mississippi, University, Mississippi 38677, U.S.A.}
\newcommand{\instUOM}{University of Miyazaki, Miyazaki 889-2192, Japan}
\newcommand{\instNovaGorica}{University of Nova Gorica, 5000 Nova Gorica, Slovenia}
\newcommand{\instPittsburgh}{University of Pittsburgh, Pittsburgh, Pennsylvania 15260, U.S.A.}
\newcommand{\instUSTC}{University of Science and Technology of China, Hefei 230026, China}
\newcommand{\instSAlabama}{University of South Alabama, Mobile, Alabama 36688, U.S.A.}
\newcommand{\instSCarolina}{University of South Carolina, Columbia, South Carolina 29208, U.S.A.}
\newcommand{\instSydney}{School of Physics, University of Sydney, New South Wales 2006, Australia}
\newcommand{\instTabuk}{Department of Physics, Faculty of Science, University of Tabuk, Tabuk 71451, Saudi Arabia}
\newcommand{\instUTokyo}{Department of Physics, University of Tokyo, Tokyo 113-0033, Japan}
\newcommand{\instIPMU}{Kavli Institute for the Physics and Mathematics of the Universe (WPI), University of Tokyo, Kashiwa 277-8583, Japan}
\newcommand{\instVictoria}{University of Victoria, Victoria, British Columbia, V8W 3P6, Canada}
\newcommand{\instVPI}{Virginia Polytechnic Institute and State University, Blacksburg, Virginia 24061, U.S.A.}
\newcommand{\instWayneState}{Wayne State University, Detroit, Michigan 48202, U.S.A.}
\newcommand{\instYamagata}{Yamagata University, Yamagata 990-8560, Japan}
\newcommand{\instYerevan}{Alikhanyan National Science Laboratory, Yerevan 0036, Armenia}
\newcommand{\instYonsei}{Yonsei University, Seoul 03722, South Korea}
\affiliation{\instSinica}
\affiliation{\instBeihang}
\affiliation{\instBUAP}
\affiliation{\instBNL}
\affiliation{\instBINP}
\affiliation{\instCMU}
\affiliation{\instCPPM}
\affiliation{\instCinvestavIPN}
\affiliation{\instPrague}
\affiliation{\instChiangMai}
\affiliation{\instChiba}
\affiliation{\instChonnam}
\affiliation{\instConacyt}
\affiliation{\instDESY}
\affiliation{\instDuke}
\affiliation{\instDuyTan}
\affiliation{\instENEA}
\affiliation{\instEri}
\affiliation{\instJuelich}
\affiliation{\instFuJen}
\affiliation{\instFudan}
\affiliation{\instGoettingen}
\affiliation{\instGifu}
\affiliation{\instSOKENDAI}
\affiliation{\instGyeongsang}
\affiliation{\instHanyang}
\affiliation{\instKEK}
\affiliation{\instJPARC}
\affiliation{\instHSE}
\affiliation{\instIISER}
\affiliation{\instIITBhubaneswar}
\affiliation{\instIITGuwahati}
\affiliation{\instIITHyderabad}
\affiliation{\instIITMadras}
\affiliation{\instIndiana}
\affiliation{\instIHEPRussia}
\affiliation{\instHEPHYVienna}
\affiliation{\instIHEPChina}
\affiliation{\instChennai}
\affiliation{\instIPP}
\affiliation{\instIOP}
\affiliation{\instIFIC}
\affiliation{\instFrascati}
\affiliation{\instNapoliINFN}
\affiliation{\instPadovaINFN}
\affiliation{\instPerugiaINFN}
\affiliation{\instPisaINFN}
\affiliation{\instRomaINFN}
\affiliation{\instRomaTreINFN}
\affiliation{\instTorinoINFN}
\affiliation{\instTriesteINFN}
\affiliation{\instJAEA}
\affiliation{\instMainz}
\affiliation{\instGiessen}
\affiliation{\instKarlsruhe}
\affiliation{\instKennesaw}
\affiliation{\instKitasato}
\affiliation{\instKISTI}
\affiliation{\instKorea}
\affiliation{\instKSU}
\affiliation{\instKyotoU}
\affiliation{\instKyungpook}
\affiliation{\instLAL}
\affiliation{\instLPI}
\affiliation{\instLNNU}
\affiliation{\instLMU}
\affiliation{\instLuther}
\affiliation{\instMNITJaipur}
\affiliation{\instMPP}
\affiliation{\instMPGHLL}
\affiliation{\instMcGill}
\affiliation{\instMETU}
\affiliation{\instMEPhI}
\affiliation{\instNagoya}
\affiliation{\instNagoyaKMI}
\affiliation{\instNaraWu}
\affiliation{\instUNAM}
\affiliation{\instNTUTaiwan}
\affiliation{\instNUUTaiwan}
\affiliation{\instKrakow}
\affiliation{\instNiigata}
\affiliation{\instNSU}
\affiliation{\instOkinawa}
\affiliation{\instOsakaCity}
\affiliation{\instRCNP}
\affiliation{\instPNNL}
\affiliation{\instPanjab}
\affiliation{\instPeking}
\affiliation{\instPanjabPAU}
\affiliation{\instRIKEN}
\affiliation{\instXavier}
\affiliation{\instSeoul}
\affiliation{\instShandong}
\affiliation{\instSPU}
\affiliation{\instSoochow}
\affiliation{\instSoongsil}
\affiliation{\instLjubljanaJSI}
\affiliation{\instKyiv}
\affiliation{\instTata}
\affiliation{\instTUM}
\affiliation{\instECUTUM}
\affiliation{\instTelAviv}
\affiliation{\instToho}
\affiliation{\instTohoku}
\affiliation{\instTitech}
\affiliation{\instTokyoMetropolitan}
\affiliation{\instUAS}
\affiliation{\instNapoliUNIV}
\affiliation{\instPadovaUNIV}
\affiliation{\instPerugiaUNIV}
\affiliation{\instPisaUNIV}
\affiliation{\instRomaUNIV}
\affiliation{\instRomaTreUNIV}
\affiliation{\instTorinoUNIV}
\affiliation{\instTriesteUNIV}
\affiliation{\instMontreal}
\affiliation{\instIPHC}
\affiliation{\instAdelaide}
\affiliation{\instBonn}
\affiliation{\instUBC}
\affiliation{\instCincinnati}
\affiliation{\instFlorida}
\affiliation{\instHamburg}
\affiliation{\instHawaii}
\affiliation{\instHeidelberg}
\affiliation{\instLjubljanaUniLJ}
\affiliation{\instLouisville}
\affiliation{\instMalaya}
\affiliation{\instLjubljanaUM}
\affiliation{\instMelbourne}
\affiliation{\instMississippi}
\affiliation{\instUOM}
\affiliation{\instNovaGorica}
\affiliation{\instPittsburgh}
\affiliation{\instUSTC}
\affiliation{\instSAlabama}
\affiliation{\instSCarolina}
\affiliation{\instSydney}
\affiliation{\instTabuk}
\affiliation{\instUTokyo}
\affiliation{\instIPMU}
\affiliation{\instVictoria}
\affiliation{\instVPI}
\affiliation{\instWayneState}
\affiliation{\instYamagata}
\affiliation{\instYerevan}
\affiliation{\instYonsei}
  \author{F.~Abudin{\'e}n}\affiliation{\instTriesteINFN} 
  \author{I.~Adachi}\affiliation{\instKEK}\affiliation{\instSOKENDAI} 
  \author{R.~Adak}\affiliation{\instFudan} 
  \author{K.~Adamczyk}\affiliation{\instKrakow} 
  \author{P.~Ahlburg}\affiliation{\instBonn} 
  \author{J.~K.~Ahn}\affiliation{\instKorea} 
  \author{H.~Aihara}\affiliation{\instUTokyo} 
  \author{N.~Akopov}\affiliation{\instYerevan} 
  \author{A.~Aloisio}\affiliation{\instNapoliUNIV}\affiliation{\instNapoliINFN} 
  \author{F.~Ameli}\affiliation{\instRomaINFN} 
% ...
\collaboration{Belle II Collaboration}


\begin{abstract}
We report on the reconstruction of various charmless $B$ decays from electron-positron collisions at the energy corresponding to the $\Upsilon(4S)$ resonance collected with the Belle II detector at the SuperKEKB collider. We use simulation to devise optimized event selections and apply them to the full data set collected in 2019,  corresponding to 8.7\,fb$^{-1}$ of integrated luminosity. We fit the difference between half of the collision energy and the $B$ candidate energy (in the $\Upsilon(4S)$ frame) for events restricted to a signal-rich range in beam-energy-constrained mass to search for charmless signals. Signal yields of approximately 80, 15, 20, 30, 90, and 160 decays are reconstructed for the channels $B^0 \to K^+\pi^-$, $B^0 \to \pi^+\pi^-$, $B^+ \to K^0_S(\to \pi^+\pi^-)\pi^+$, $B^+ \to K^+\pi^0(\to \gamma\gamma)$, $B^+ \to K^+K^-K^+$,  and $B^+ \to K^+\pi^-\pi^+$, respectively. Yields and background contaminations are compatible with those expected from simulation and comparable with those obtained by the Belle experiment.
The results show a good understanding of the detector performance and offer a reliable basis to assess projections for future reach.

\end{abstract}


\maketitle

{\renewcommand{\thefootnote}{\fnsymbol{footnote}}}
\setcounter{footnote}{0}



\section{Introduction and motivation}

The study of charmless $B$ decays is a keystone of the Belle~II physics program, which offers the unique capability of studying jointly, within a consistent experimental environment, all relevant two-, three-, and multi-body final states. This ability can enable significant advances, including an improved determination of the quark-mixing-matrix angle $\phi_2/\alpha$,  a conclusive understanding of long-standing anomalies like the so-called $K\pi$ puzzle, and a thorough investigation of charge-parity-violating asymmetries localized in the phase space of three-body $B$ decays. Ultimately, the Belle II charmless program is expected to offer a sharper picture of the weak interactions of quarks that could reveal possible deviations from the standard model~\cite{Kou:2018nap}. \par 
The Belle~II detector, complete with its vertex detector, started its 2019 collision operations on March 11 and continued until December 13.  The collected sample of electron-positron collisions corresponds to an integrated luminosity of 8.7\,fb$^{-1}$ at the $\Upsilon(4{\rm S})$ resonance and 0.827\,fb$^{-1}$ at an energy about 60\,MeV smaller.  This \mbox{document} describes the reconstruction of  charmless signals in addition to the 25 \mbox{$B^0 \to K^+ \pi^-$} decays reconstructed in Fall 2019 using half of the current data set~\cite{Benedikt:2019}. 

We focus on two- and three-body charmless decays with branching fractions of $10^{-6}$, or larger, into final states sufficiently simple to obtain visible signals in the current data set with a relatively straightforward reconstruction.  The target decay modes are  \mbox{$B^0 \to K^+ \pi^-$}, \mbox{$B^0 \to \pi^+\pi^-$}, \mbox{$B^+ \to K^0_S(\to \pi^+\pi^-)\pi^+$}, \mbox{$B^+ \to K^+\pi^0 (\to \gamma\gamma)$}, \mbox{$B^+ \to K^+K^-K^+$},  and \mbox{$B^+ \to K^+\pi^-\pi^+$}.
Charge-conjugate processes are implied in what follows.\par
The reconstruction strategy and procedures are developed and finalized in simulated data prior to applying it to the experimental data. Experimental and simulated data are then compared in terms of signal yields, backgrounds, and relevant resolutions.
Most of the analysis is conducted using the following variables, which are known to be strongly discriminating between $B$ signal and background from $e^+e^- \to q\bar{q}$ continuum events, where $q$ indicates any quark of the first or second family:
\begin{itemize}
    \item the energy difference $\Delta E \equiv \sqrt{s}/2 - E^{*}_{B}$ between half of the collision energy and the total energy of the reconstructed $B$ candidate, both in the $\Upsilon(4S)$ frame;
    \item the beam-energy-constrained mass $M_{\rm bc} \equiv \sqrt{s/(4c^4) - (p^{*}_B/c)^2}$, which is the invariant mass of the $B$ candidate where the $B$ energy is replaced by the (more precisely known) half of the collision energy. 
\end{itemize}

\section{The Belle II detector}
Belle II is a particle-physics detector~\cite{Kou:2018nap, Abe:2010sj}, designed to reconstruct the products of electron-positron collisions produced by the SuperKEKB asymmetric-energy collider~\cite{Akai:2018mbz}, located at the KEK laboratory in Tsukuba, Japan. Belle II comprises several subdetectors arranged around the interaction space-point in a cylindrical geometry. The innermost subdetector is the vertex detector, which uses position-sensitive silicon layers to sample the trajectories of charged particles (tracks) in the vicinity of the interaction region to extrapolate the decay positions of their long-lived parent particles. The vertex detector includes two inner layers of silicon pixel sensors and four outer layers of silicon microstrip sensors. The second pixel layer is currently incomplete and covers only a small portion of azimuthal angle. Charged-particle momenta and charges are measured by a large-radius, helium-ethane, small-cell central drift chamber, which also offers charged-particle-identification information through a measurement of particles' energy-loss by specific ionization. A Cherenkov-light angle and time-of-propagation detector sourrounding the chamber provides charged-particle identification in the central detector volume, supplemented by proximity-focusing, aerogel, ring-imaging Cherenkov detectors in the forward regions. A CsI(Tl)-crystal electromagnetic calorimeter allows for energy measurements of electrons and photons.  A solenoid surrounding the calorimeter generates a uniform axial 1.5\,T magnetic field filling its inner volume. Layers of plastic scintillator and resistive-plate chambers, interspersed between the
magnetic flux-return iron plates, allow for identification of $K^0_{\rm L}$ and muons.
The subdetectors most relevant for this work are the silicon vertex detector, the tracking drift chamber, the particle-identification detectors, and the electromagnetic calorimeter.

\section{Selection and reconstruction}
\label{sec:selection}

We reconstruct the following two-body decays:

\begin{itemize}
    \item $B^0 \to K^+\, \pi^-$,
    \item $B^0 \to \pi^+\, \pi^-$,
        \item $B^+ \to K_{\rm S}^0(\to \pi^+\pi^-)\, \pi^+$,
    \item $B^+ \to K^+\, \pi^0 (\to \gamma\gamma)$,
\end{itemize}
and three-body decays:
\begin{itemize}
    \item $B^+ \to K^+\, K^-\, K^+$,
    \item $B^+ \to K^+\, \pi^+\, \pi^-$.
\end{itemize}
In addition, we use the control channel \mbox{$B^+ \to \overline{D}^0 (\to K^+ \pi^- \pi^0)\, \pi^+$} for validation of continuum-suppression discriminating variables and optimization of the $\pi^0$ selection.


\subsection{Data}
We use generic simulated data to optimize the event selection and compare the final $\Delta E$ distributions observed in experimental data with expectations. We use signal-only simulated data to model relevant signal features for fits.
Generic simulation consists of Monte Carlo samples that include $\BzBzb$, $\BpBm$, $\uubar$, $\ddbar$, $\ccbar$, and $\ssbar$ processes in realistic proportions and correspond to an integrated luminosity of 50\,fb$^{-1}$, about six times the $\Upsilon$(4S) data. In addition, $2\times 10^6$ signal-only events are generated for each channel~\cite{Ryd:2005zz}.
Three-body decays are generated assuming phase-space distributions. \par 
As for experimental data, we use all 2019  $\Upsilon$(4S) good-quality runs, corresponding to an integrated luminosity of 8.7\,fb$^{-1}$. All events are required to meet mild data-skim selection criteria, based on total energy and charged-particle multiplicity in the event, targeted at reducing sample sizes to a manageable level. All data are processed using the Belle~II analysis software framework~\cite{Kuhr:2018lps}.

\subsection{Reconstruction and baseline selection}
We form final-state particle candidates by  applying loose baseline selection criteria and then combine them in kinematic fits consistent with the topologies of the desired decays to reconstruct intermediate states and $B$ candidates. \par We reconstruct charged pion and kaon candidates by starting from the most inclusive charged-particle classes  and by requiring fiducial criteria that restrict them to the full acceptance in the central drift chamber and to loose ranges in impact parameter to reduce beam-background-induced tracks, which do not originate from the interaction region preferably.
We reconstruct neutral pion candidates by requiring photons to exceed energies of about $20$\,MeV, restricting the diphoton mass, and excluding extreme helicity-angle values  to suppress combinatorial background from collinear soft photons. The mass of the $\pi^0$ candidates is constrained to its known value in subsequent kinematic fits. 
For $K_{\rm S}^0$ reconstruction, we use pairs of opposite-charge red particles that originate from a common space-point and have dipion mass consistent with a $K_{\rm S}^0$. The resulting $K^\pm$, $\pi^\pm$, $\pi^0$, and $\PKzS$ candidates are combined through kinematic fits into each of our target signal channels, consistent with the desired topology. Because we use flavor-tagging information as input to the continuum-background discriminator, we reconstruct the vertex of the accompanying tag-side $B$ mesons using  all tracks in the tag-side and identify the flavor using a category-based flavor tagger~\cite{Abudinen:2018}. 
The reconstruction of the control channel is conceptually similar, except for the requirement \mbox{$1.84 < m(K^-\pi^+\pi^0)< 1.89$\,GeV/$c^2$} on the  $K^-\pi^+\pi^0$ mass to be consistent with the known $D^0$ meson mass. 
\par
Simulation is used to identify and suppress contamination from peaking backgrounds, that is, misreconstructed events clustering in the signal region $M_{\rm bc} > 5.27$\,GeV/$c^2$ and \mbox{$-0.15 < \Delta E < 0.15$ GeV}. 
Relevant peaking backgrounds  affect only the \mbox{$B^+ \to K^+\pi^-\pi^+$} channel. Background from \mbox{$B^+ \to \overline{D}^0(\to K^+\pi^-)\pi^+$} decays is suppressed by vetoing candidates with kaon-pion mass \mbox{$1.84 < m(K^+\pi^-) < 1.89$\,GeV/$c^2$}.  The contribution from \mbox{$B^+ \to J/\psi(\to \mu^+\mu^-)K^+$} decays where muons are misidentified as pions is suppressed by vetoing candidates with dipion mass \mbox{$3.05 < m(\pi^+\pi^-) < 3.15$\,GeV/$c^2$}.

\subsection{Continuum suppression}
The main challenge in reconstructing significant charmless signals is the large contamination from continuum background. To discriminate against such background, we use a binary boosted decision-tree classifier that combines nonlinearly a number of variables known to provide statistical discrimination between $B$-meson signals and continuum.   We choose 39 variables whose correlation with $\Delta E$ and $M_{\it bc}$ is below $\pm$5\% to avoid biases in signal-yield determination. These variables include quantities associated to event topology (global and signal-only angular configurations), flavor-tagger information, vertex separation and uncertainty information, and kinematic-fit quality information. 
Data-simulation comparison for input distributions using the control sample shows no major inconsistency for both signal and background.
We train the classifier to identify statistically significant signal and background features using unbiased simulated samples. 


\section{Optimization of the signal selection}
\label{sec:yields}
For each channel, we optimize the selection to isolate abundant, low-background signals using simulated and control-sample data. We vary the selection criteria on continuum-suppression output, charged-particle identification information, and choice of $\pi^0$ (when appropriate) to maximize ${\rm S}/\sqrt{{\rm S}+{\rm B}}$, where ${\rm S}$ and ${\rm B}$ are signal and background yields, respectively, estimated in the $\Delta E$ signal region. Continuum-suppression and particle-identification requirements are optimized simultaneously using simulated data. The $\pi^0$ selection is optimized independently by using control \mbox{$B^+ \to \overline{D}^0(\to K^+\pi^-\pi^0)\pi^-$} decays in which S is the \mbox{$B^+ \to \overline{D}^0(\to K^+\pi^-\pi^0)\pi^-$} signal yield, scaled to the expected \mbox{$B^+ \to K^+\pi^0$} yield, and B is the background observed in an $M_{\rm bc}$ sideband of \mbox{$B^+ \to K^+\pi^0$}.


\section{Determination of signal yields}
\label{sec:yields}
More than one candidate per event populates the resulting $\Delta E$ distributions, with average multiplicities ranging from 1.00 to 1.25. We restrict to one candidate per event as follows. For channels with $\pi^0$, we first select the $\pi^0$ candidate with the highest $p$-value of the mass-constrained diphoton fit. If more than one candidate remains, and for all other channels, we select a single $B$ candidate randomly. 

Signal yields are determined with maximum likelihood fits of the unbinned $\Delta E$ distributions of candidates restricted to the signal region in $M_{\rm bc}$. Fit models are generally determined empirically by using simulation, with the only additional flexibility of a global shift of peak positions when suggested by likelihood-ratio tests.

\section{Results}

Figures~1--6 show the resulting $\Delta E$ distributions, with fit results overlaid.
Prominent narrow signals of 10--150 events are visible overlapping smooth backgrounds dominated by continuum. The $B^0 \to K^+\pi^0$ signal shows a low-$\Delta E$ tail, due to resolution effects in $\pi^0$ reconstruction.
Satellite signals from kinematically similar misreconstructed decays are visible in the $B^0 \to K^+\pi^- $, $B^0 \to \pi^+\pi^-$, and $B^+ \to K^+\pi^-\pi^-$ decays.
\begin{figure}[htb]
 \centering
 \includegraphics[width=0.475\textwidth]{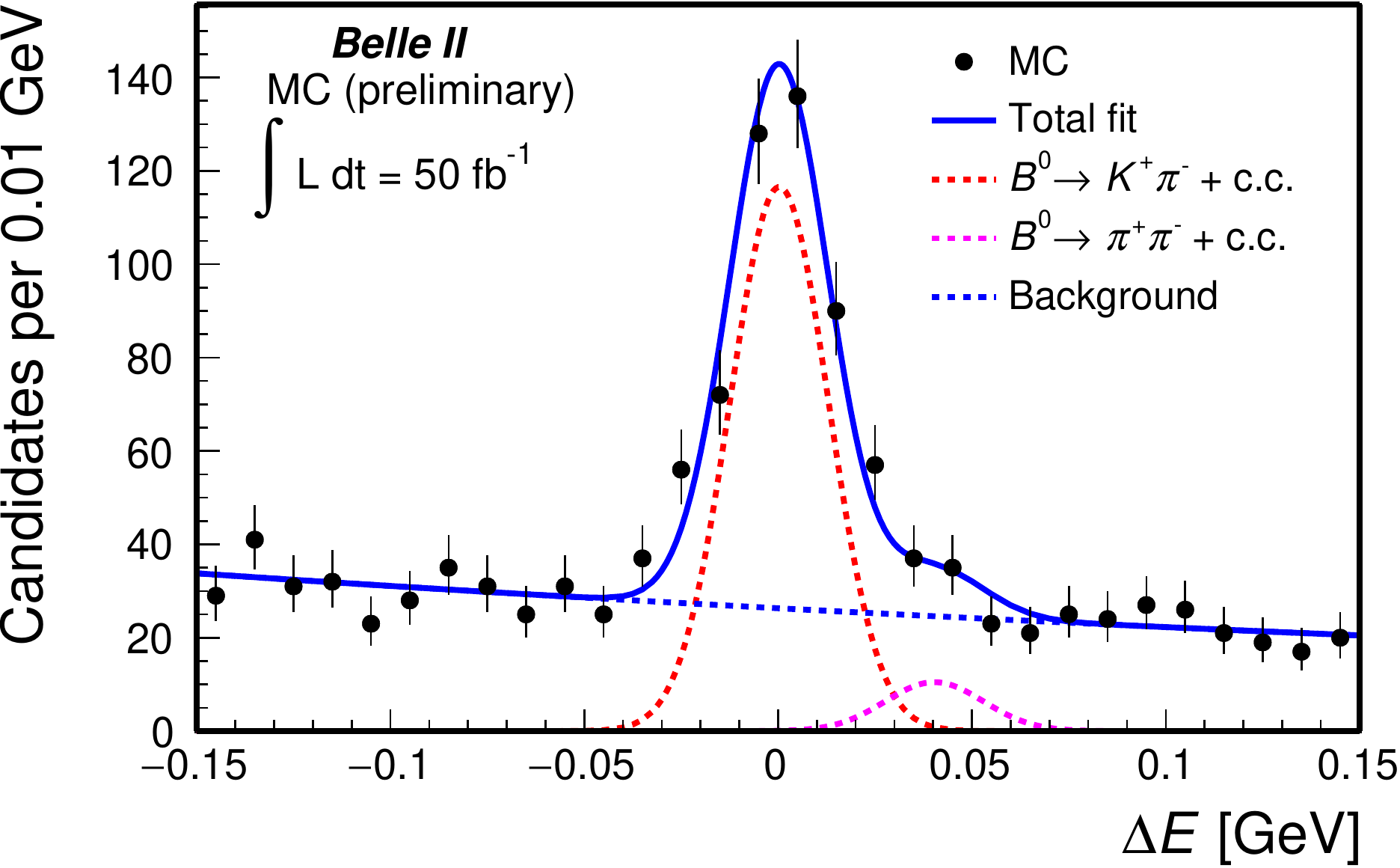}
 \includegraphics[width=0.475\textwidth]{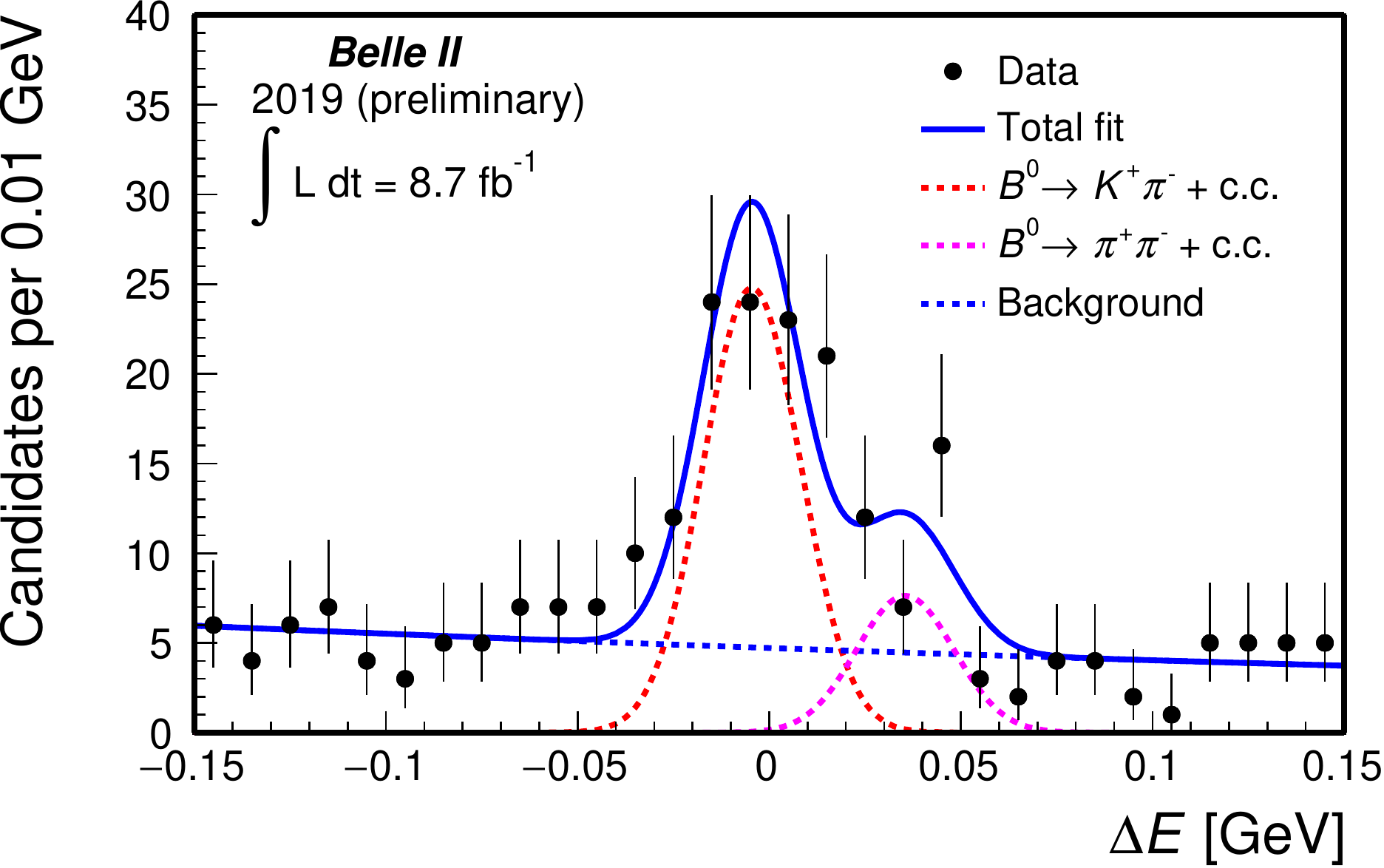}
 \caption{Distributions of $\Delta E$ for $B^0 \to K^+\pi^-$  candidates reconstructed in (left) simulated data and (right) 2019 Belle II data selected through the baseline criteria plus an optimized continuum-suppression and kaon-enriching selection, and further restricted to $M_{\rm bc} > 5.27$\,GeV/$c^2$. A misreconstructed $\pi^+\pi^-$~component is included with shape equal to the $K^+\pi^-$ shape and distance from the $K^+\pi^-$ peak fixed to the known value. The global position of the two peaks is determined by the fit. The projection of an unbinned maximum likelihood fit is overlaid.}
 \label{fig:fit_dE_Kpi_MC_rand}
\end{figure}

\begin{figure}[htb]
 \centering
 \includegraphics[width=0.475\textwidth]{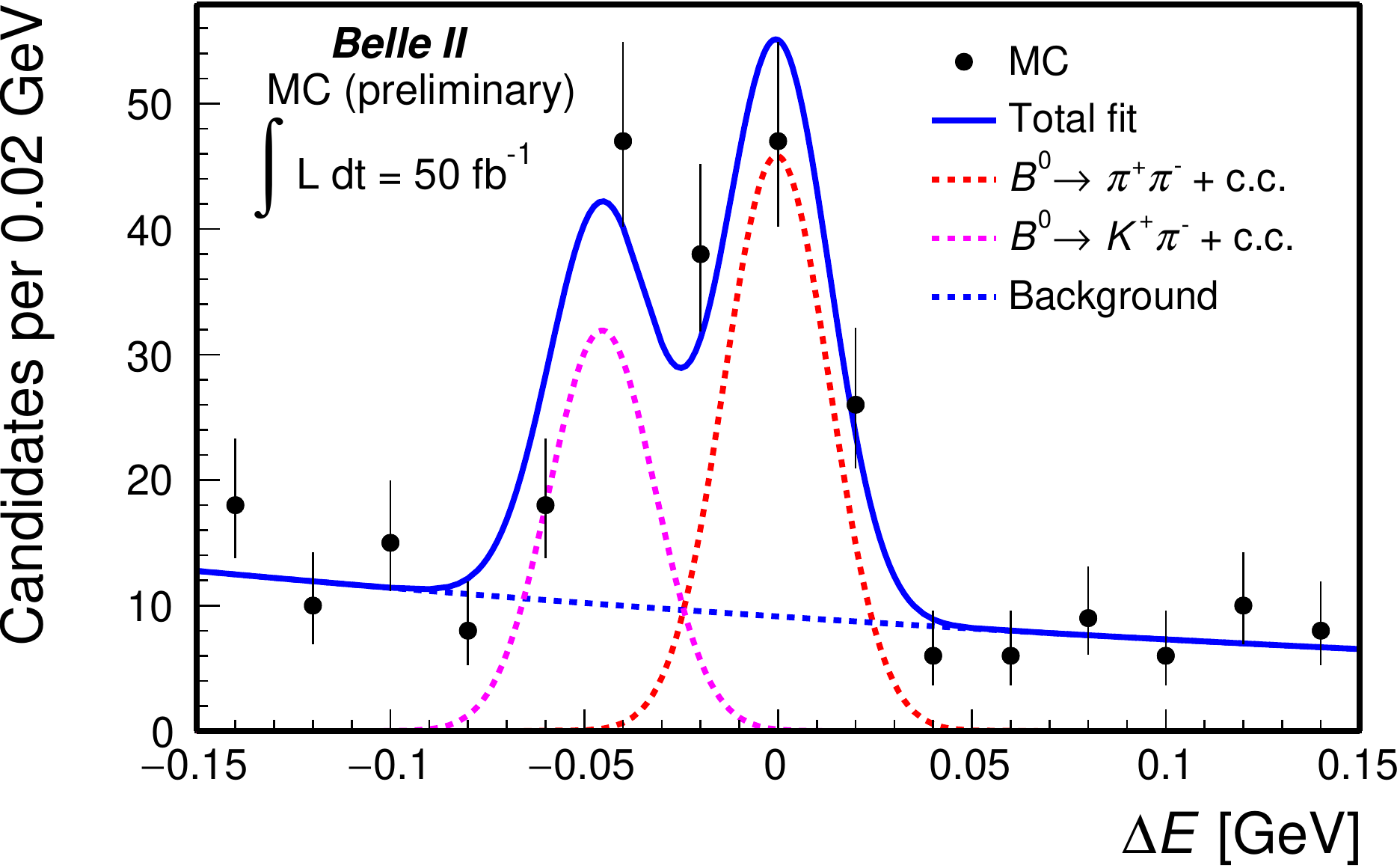}
 \includegraphics[width=0.475\textwidth]{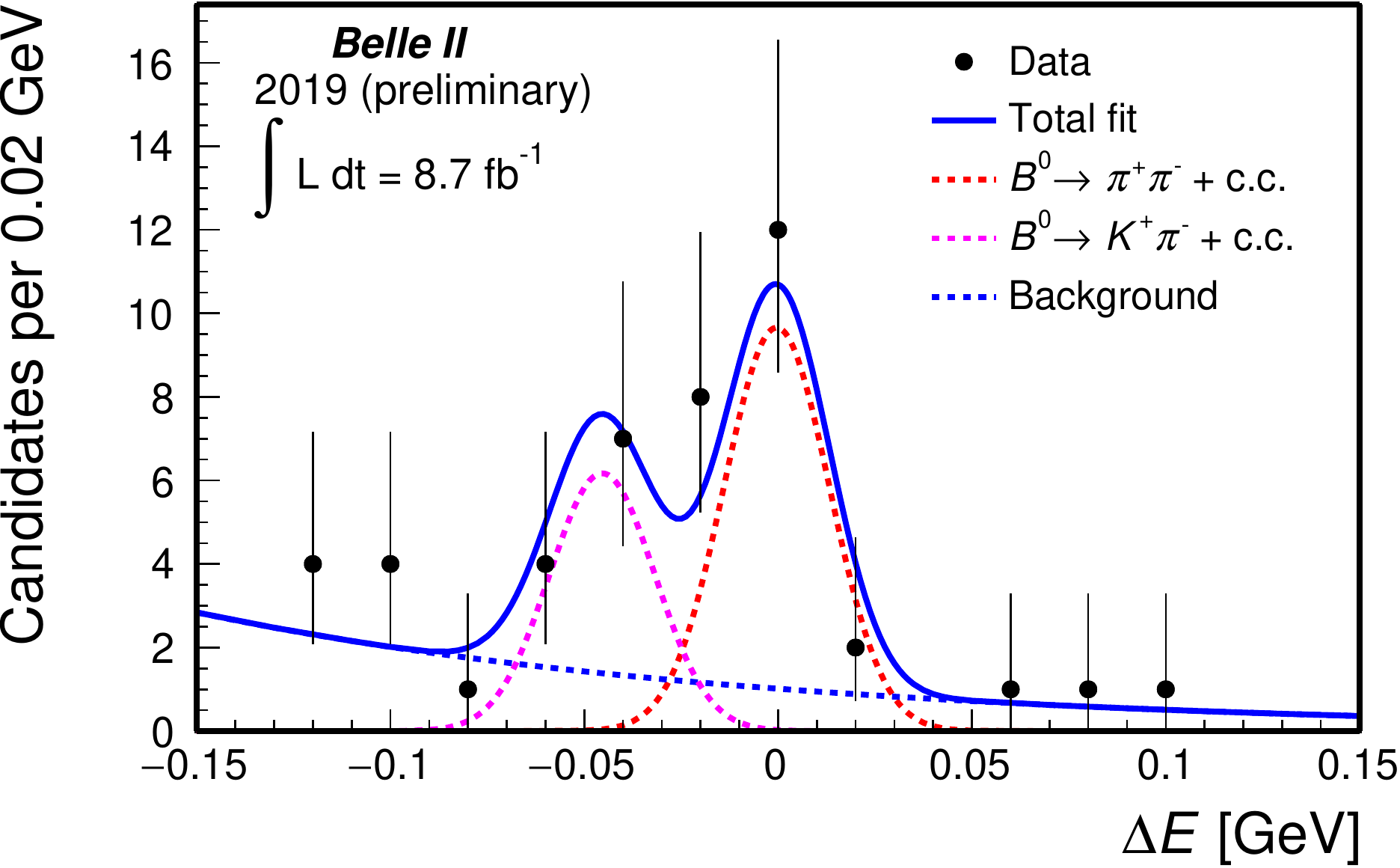}
 \caption{Distributions of $\Delta E$ for $B^0 \to \pi^+\pi^-$  candidates reconstructed in (left) simulated data and (right) 2019 Belle II  data,  selected through the baseline criteria plus an optimized continuum-suppression and pion-enriching selection, and further restricted to $M_{\rm bc} > 5.27$\,GeV/$c^2$.  A misreconstructed $K^+\pi^-$~component is included with shape equal to the $\pi^+\pi^-$ shape and distance from the $\pi^+\pi^-$ peak fixed to the known value. The global position of the two peaks is determined by the fit.
 The projection of an unbinned maximum likelihood fit is overlaid.}
 \label{fig:fit_dE_pipi_MC_rand}
\end{figure}

\begin{figure}[htb]
 \centering
 \includegraphics[width=0.475\textwidth]{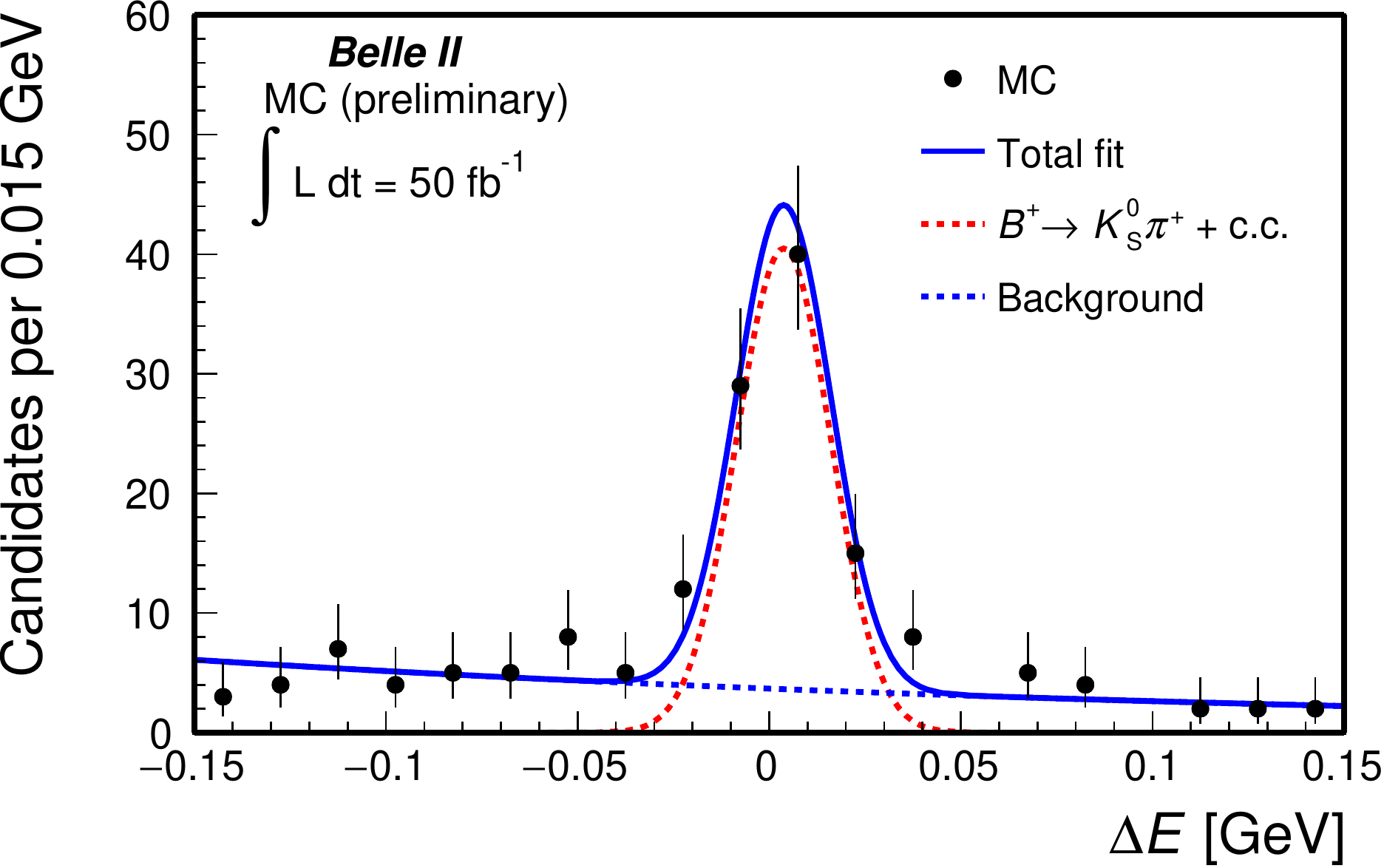}
 \includegraphics[width=0.475\textwidth]{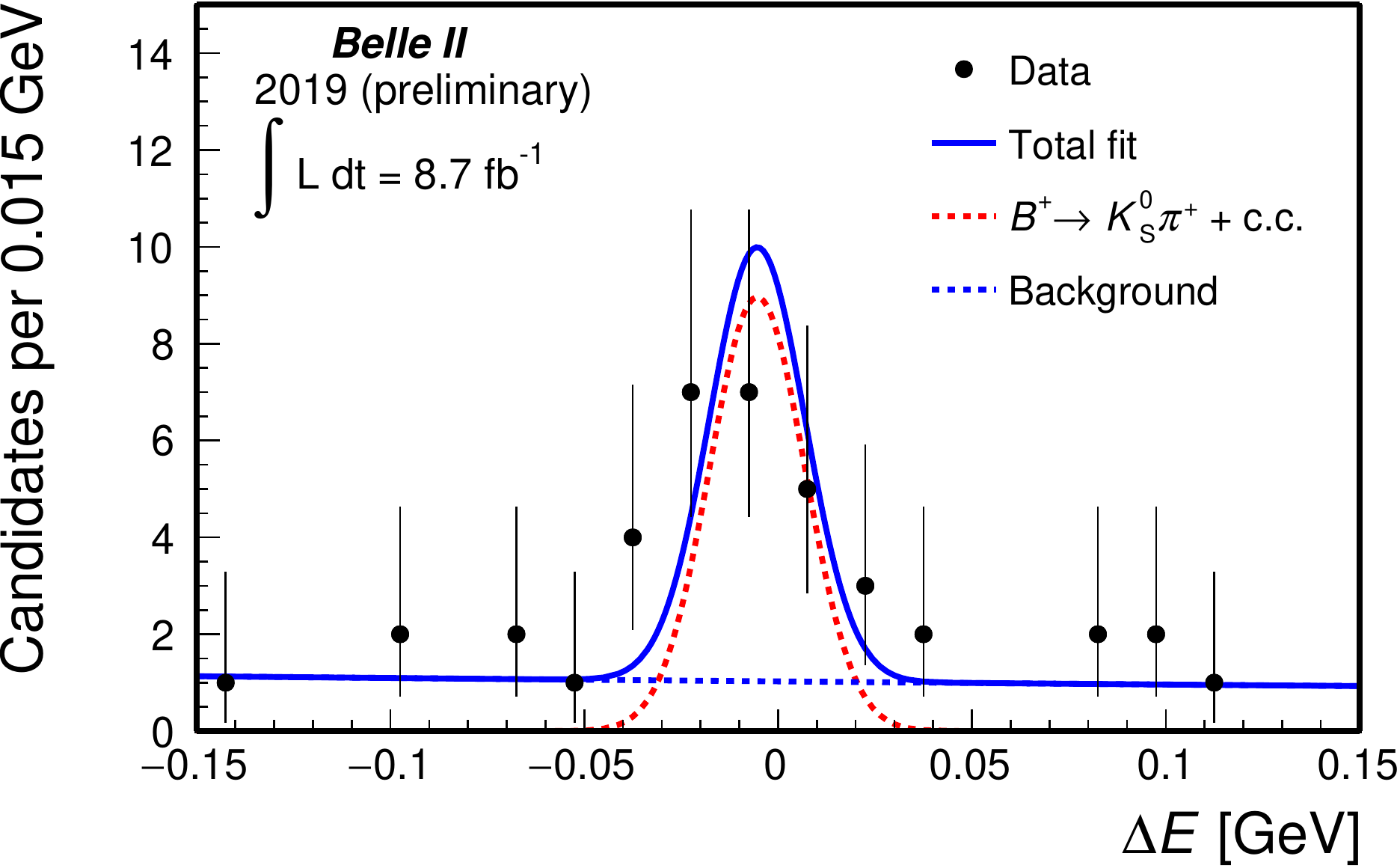}
 \caption{Distributions of $\Delta E$ for $B^+ \to \PKzS \pi^+$  candidates reconstructed in (left) simulated data and (right) 2019 Belle II data, selected through the baseline criteria plus an optimized continuum-suppression, and further restricted to $M_{\rm bc} > 5.27$\,GeV/$c^2$. The projection of an unbinned maximum likelihood fit is overlaid.}
 \label{fig:fit_dE_KSpi_MC_rand}
\end{figure}

\begin{figure}[htb]
 \centering
 \includegraphics[width=0.475\textwidth]{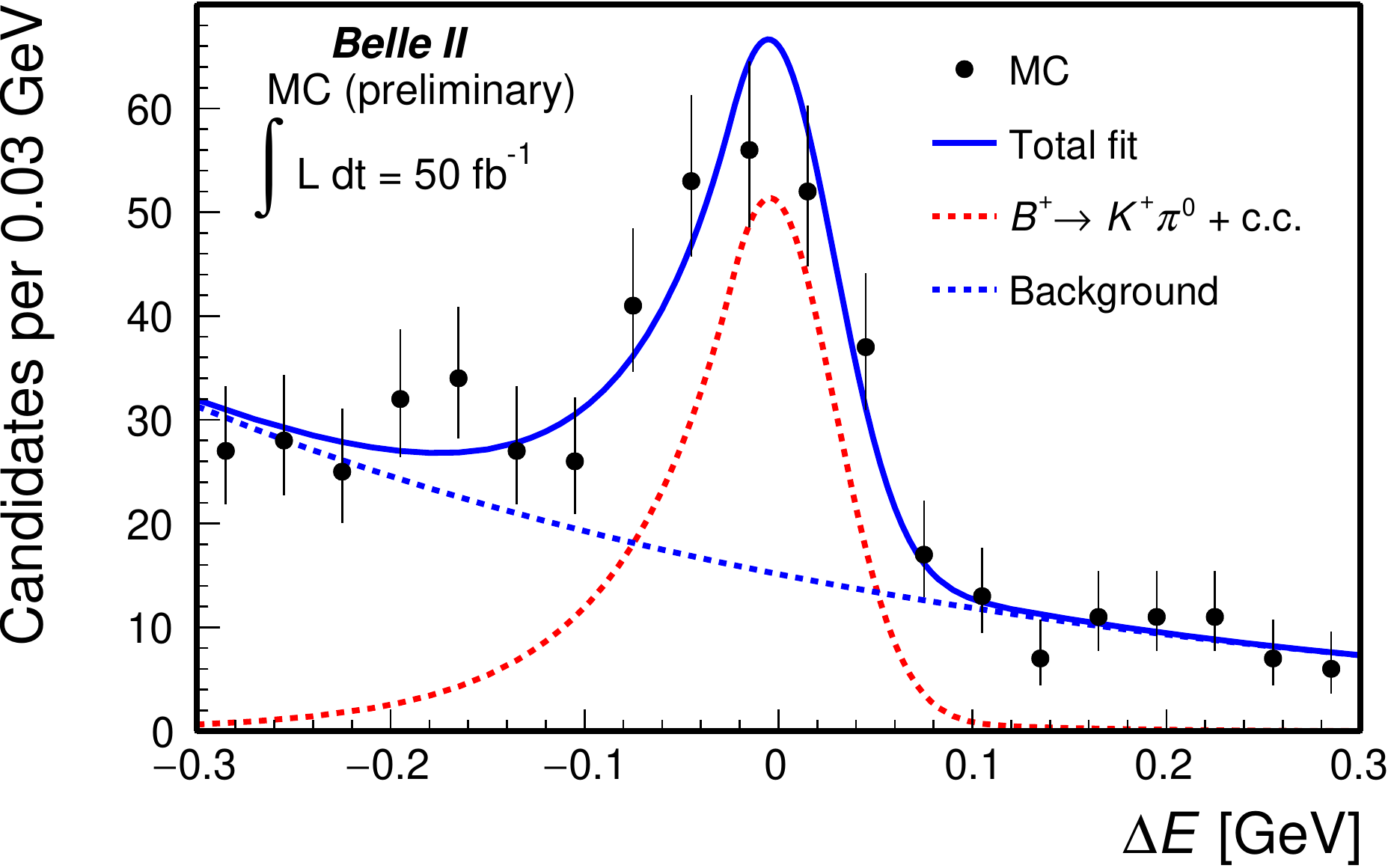}
 \includegraphics[width=0.475\textwidth]{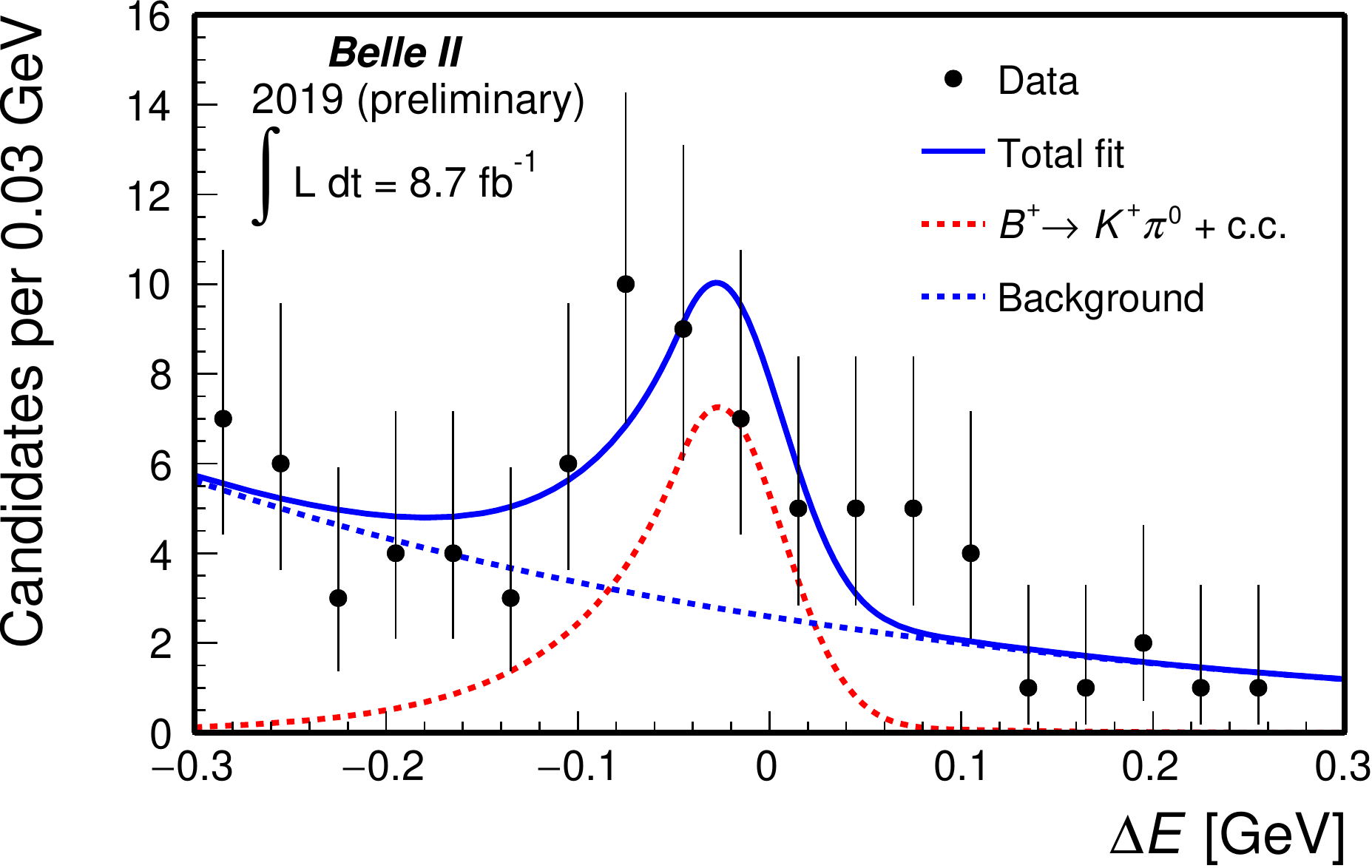}
 \caption{Distributions of $\Delta E$ for $B^+ \to K^+\pi^{0}$  candidates reconstructed in (left) simulated data and (right) 2019 Belle II data, selected through the baseline criteria plus an optimized continuum-suppression, kaon-  and $\pi^0$-enriching  selection, further restricted to $M_{\rm bc} > 5.27$\,GeV/$c^2$. The projection of an unbinned maximum likelihood fit is overlaid.}
 \label{fig:fit_dE_Kpi0_MC_rand}
\end{figure}

\begin{figure}[htb]
 \centering
 \includegraphics[width=0.475\textwidth]{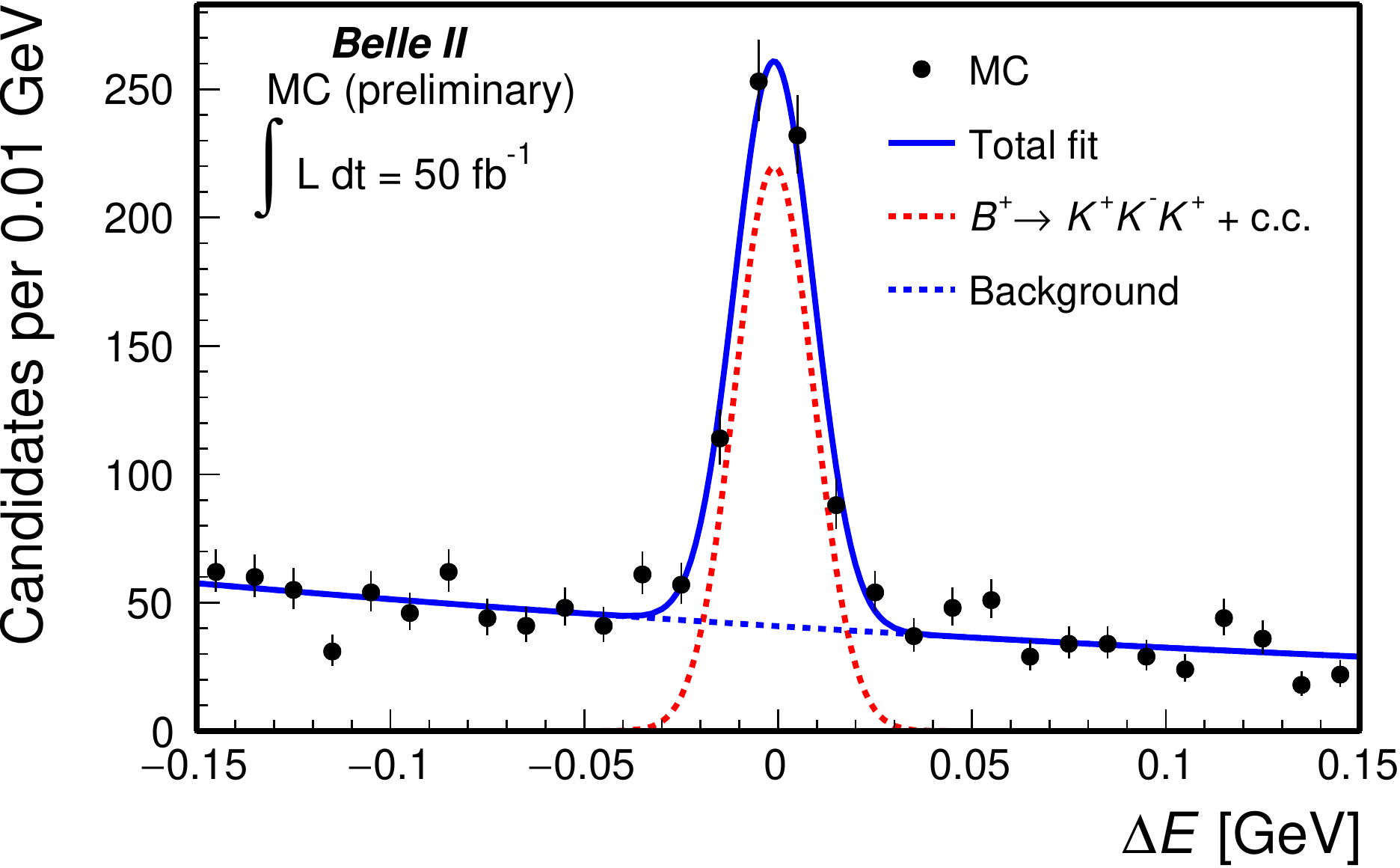}
 \includegraphics[width=0.475\textwidth]{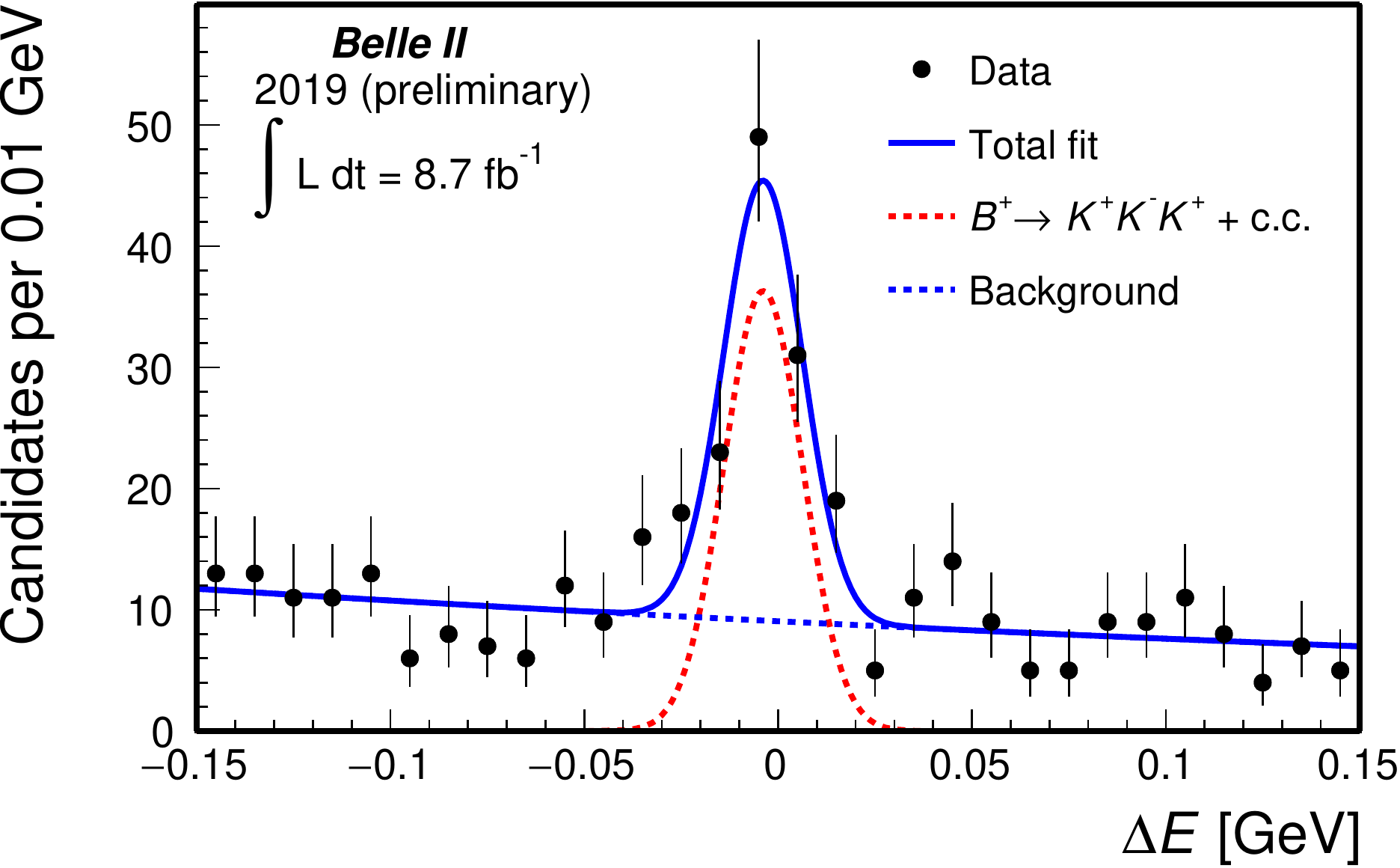}
 \caption{Distributions of $\Delta E$ for $B^+ \to K^+K^-K^+$  candidates reconstructed in (left) simulated data and (right) 2019 Belle II data, selected through the baseline criteria plus an optimized continuum-suppression and kaon-enriching selection, further restricted to $M_{\rm bc} > 5.27$\,GeV/$c^2$.  The projection of an unbinned maximum likelihood fit is overlaid.}
 \label{fig:fit_dE_KKK_MC_rand}
\end{figure}


\begin{figure}[htb]
 \centering
 \includegraphics[width=0.475\textwidth]{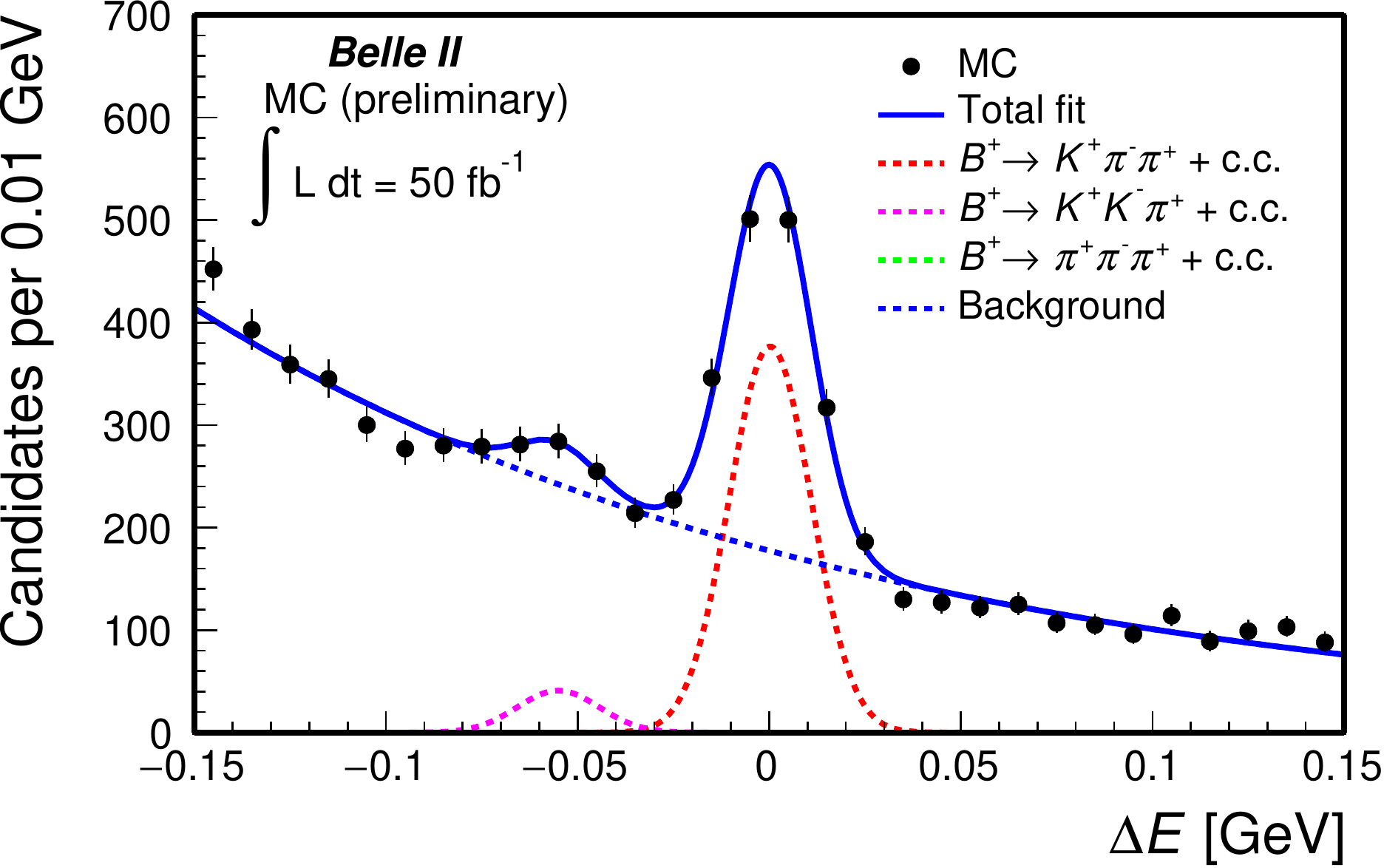}
 \includegraphics[width=0.475\textwidth]{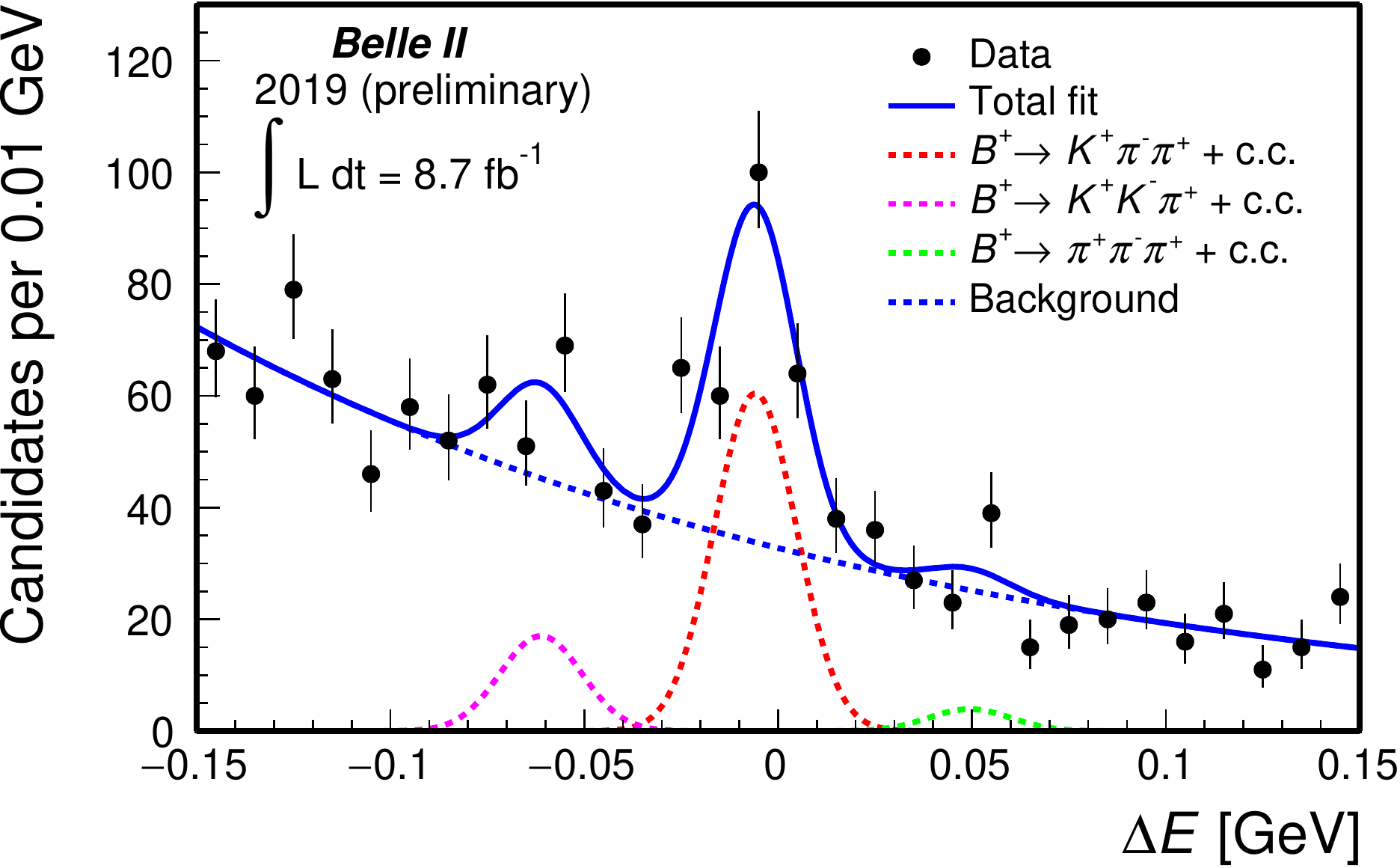}
 \caption{Distributions of $\Delta E$ for $B^+ \to K^+\pi^-\pi^+$  candidates reconstructed in (left) simulated data and (right) 2019 Belle II data, selected through the baseline criteria plus an optimized continuum-suppression and kaon-enriching selection, further restricted to $M_{\rm bc} > 5.27$\,GeV/$c^2$. Vetoes for peaking backgrounds are applied.
 Misreconstructed  $K^+K^-\pi^+$ and $\pi^+\pi^-\pi^+$ components are included with shape equal to the $K^+\pi^-\pi^+$ shape and distances from the $K^+\pi^-\pi^+$ peak fixed to the known values.  The global position of the three peaks is determined by the fit. The projection of an unbinned maximum likelihood fit is overlaid.}
 \label{fig:fit_dE_Kpipi_MC_rand}
\end{figure}

\begin{table}[!ht]
    \centering
\begin{tabular}{l  r  r  r  r }
\hline\hline
\multicolumn{1}{c}{} & 
\multicolumn{2}{c}{Yield} &   
\multicolumn{2}{c}{Yield/$\si{fb^{-1}}$} \\
Decay & \multicolumn{1}{c}{MC} & \multicolumn{1}{c}{Data} & \multicolumn{1}{c}{MC} & \multicolumn{1}{c}{Data} \\\hline
  $\PBzero\to\PKp\Pgpm$    &	$371 \pm 24$ &	$79\pm 11$ &	$7.4\pm 0.5$ &	$9.1\pm1.3$ \\ 
 $\PBzero\to\Pgpp\Pgpm$	   &   $78\pm 11$   &	$16\pm 5\enskip$ &	$1.6\pm 0.2$& 	$1.8\pm 0.6$ \\
 $\PBplus\to\PKzS\Pgpp$	   &  $83\pm 10$	& $18\pm 5\enskip$	& $1.7\pm 0.2$	& $2.1\pm 0.6$ \\
 $\PBplus\to\PKp\Pgpz$	   &   $191\pm 20$	& $27\pm 8\enskip$	&$3.8 \pm 0.4$	& $3.1\pm 0.9$ \\
 $\PBplus\to\PKp\PKp\PKm$  &  $559\pm 28 $	& $92\pm 12$	& $11.2\pm 0.6$ &	$10.6\pm 1.4$\\
 $\PBplus\to\PKp\Pgpp\Pgpm$&  $1008\pm 44$	& $160\pm 19$	& $20.2\pm 0.9$	& $18.4\pm 2.2$\\
 \hline
\end{tabular}
    \caption{Summary of charmless yields, and yields per integrated luminosity, in 2019 Belle II data. The size of the simulated (experimental) sample corresponds to an integrated luminosity of 50~(8.7)~fb$^{-1}$. Only the statistical contributions to the uncertainties are reported.} 
    \label{tab:OneCandSummary}
\end{table}{}
\clearpage
\section{Comparison with Belle}
Comparison of the current two-body results with Belle's latest results on the same channels, based on the full sample corresponding to 712\,fb$^{-1}$~\cite{Duh:2012ie}, provides interesting insight to assess Belle II's current and projected performance. A consistent comparison would require redoing the full analysis of Belle data to account for the differences in the statistical content of the variables and in analysis strategy.  We offer a simplified comparison based on signal yields and peak purities (i.e., S/B at peak), shown in Table~\ref{tab:Belle2-vs-Belle}.
The current Belle II performance in charmless $B$ decay reconstruction is comparable to the Belle performance.
\begin{table}[!htb]
\begin{tabular}{l  c  c  c  c}
\hline \hline
Decay  & Belle II yield/fb$^{-1}$ & Belle II purity & Belle yield/fb$^{-1}$ & Belle purity\\
\hline 
$B^0 \to K^+\pi^-$ & 9.1 $\pm$ 1.3 & $\approx 10$ &   10.6 $\pm$ 0.18 & $\approx 5$ \\

$B^0 \to \pi^+\pi^-$ & 1.8 $\pm$ 0.6 & $\approx 5.5$ &  2.96 $\pm$ 0.12 & $\approx 2.4$ \\

$B^+ \to K^+\pi^0$ & 3.1 $\pm$ 0.9  & $\approx 3.6$&   5.2 $\pm$ 0.13 & $\approx 3.5$ \\

$B^+ \to \PKzS\pi^+$  & 2.1 $\pm$ 0.6 & $\approx 10$ &   4.5 $\pm$ 0.1 & $\approx 12$\\



\hline\hline
\end{tabular}
\caption{Comparison between Belle~\cite{Duh:2012ie} and Belle II (this work) performance in signal yield and peak purity.
Only the statistical contributions to the uncertainties are reported.
}
\label{tab:Belle2-vs-Belle}
\end{table}

\section{Summary}
\label{sec:summary}

\begin{figure}[htb]
 \centering
 \includegraphics[width=0.7\textwidth]{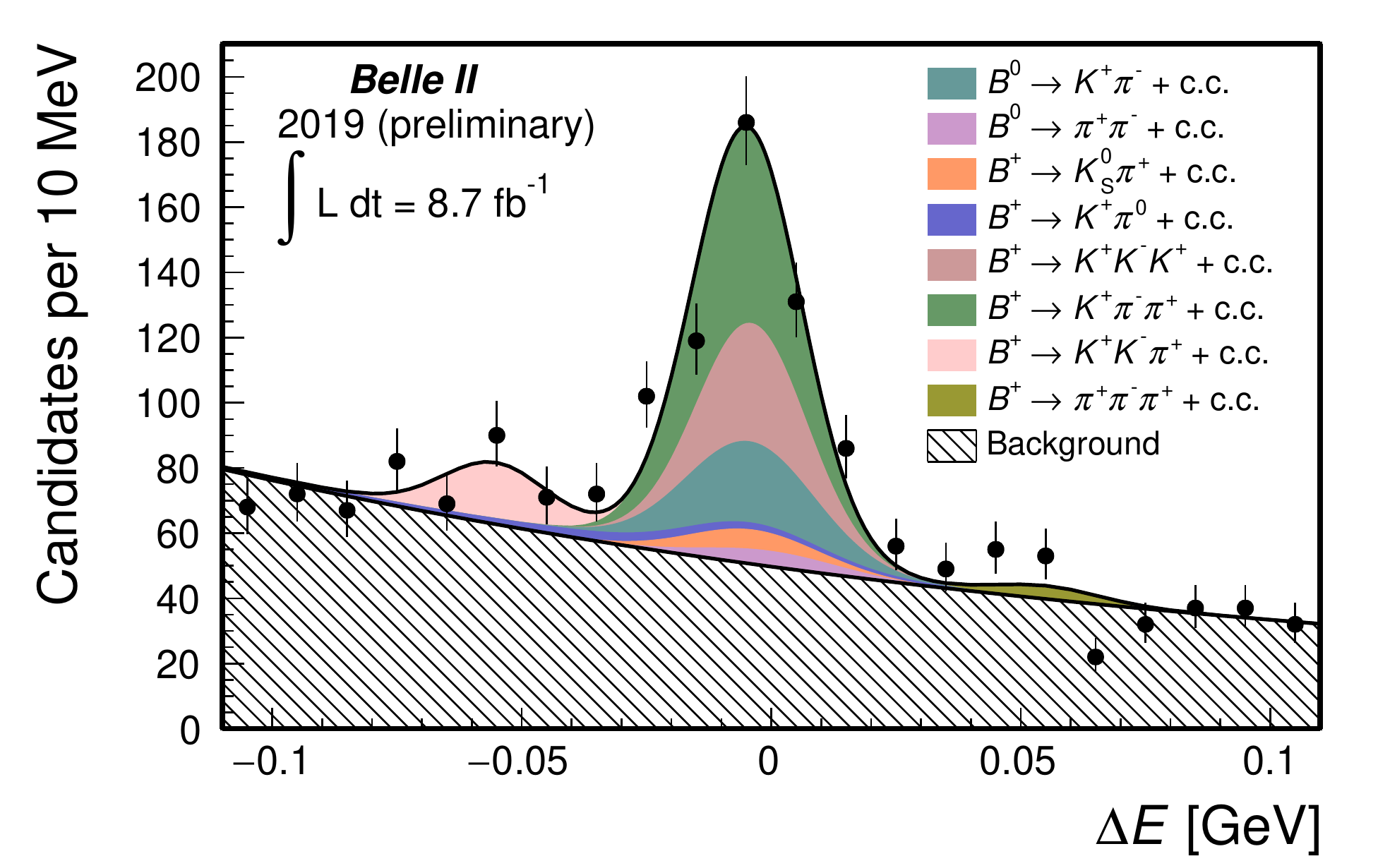}
 \caption{Stacked $\Delta E$ distributions of charmless channels reconstructed in 2019 Belle II data with the sum of fit projections overlaid.}
 \label{fig:hype}
\end{figure}

We report on the reconstruction of various $B$ charmless signals in 2019 Belle II data. We devise optimized event selections using simulation and apply them to the full data set collected in 2019,  corresponding to 8.7\,fb$^{-1}$ of integrated luminosity. The $\Delta E$ distributions of the resulting samples, restricted in $M_{\rm bc}$, are fit to search for charmless signals. Signal yields of approximately 80, 15, 20, 30, 90, and 160 decays are reconstructed for the channels \mbox{$B^0 \to K^+\pi^-$}, \mbox{$B^0 \to \pi^+\pi^-$}, \mbox{$B^+ \to K^0_S(\to \pi^+\pi^-)\pi^+$}, \mbox{$B^+ \to K^+\pi^0(\to \gamma\gamma)$}, \mbox{$B^+ \to K^+K^-K^+$},  and \mbox{$B^+ \to K^+\pi^-\pi^+$}, respectively, totaling nearly 400 charmless $B$~decays (Fig.~\ref{fig:hype}). Yields are generally compatible with expectations from simulation and have comparable backgrounds. This work establishes a good understanding of detector performance. In addition,  it establishes solid ground to assess future projections for charmless physics measurements, such as stringent tests of the isospin sum-rule~\cite{Kou:2018nap}, for which competitive results could be available with the sample collected a year from now.

\clearpage

\section*{Acknowledgments}

We thank the SuperKEKB group for the excellent operation of the
accelerator; the KEK cryogenics group for the efficient
operation of the solenoid; and the KEK computer group for
on-site computing support.
This work was supported by the following funding sources:
Science Committee of the Republic of Armenia Grant No. 18T-1C180;
Australian Research Council and research grant Nos.
DP180102629, 
DP170102389, 
DP170102204, 
DP150103061, 
FT130100303, 
and
FT130100018; 
Austrian Federal Ministry of Education, Science and Research, and
Austrian Science Fund No. P 31361-N36; 
Natural Sciences and Engineering Research Council of Canada, Compute Canada and CANARIE;
Chinese Academy of Sciences and research grant No. QYZDJ-SSW-SLH011,
National Natural Science Foundation of China and research grant Nos.
11521505,
11575017,
11675166,
11761141009,
11705209,
and
11975076,
LiaoNing Revitalization Talents Program under contract No. XLYC1807135,
Shanghai Municipal Science and Technology Committee under contract No. 19ZR1403000,
Shanghai Pujiang Program under Grant No. 18PJ1401000,
and the CAS Center for Excellence in Particle Physics (CCEPP);
the Ministry of Education, Youth and Sports of the Czech Republic under Contract No.~LTT17020 and 
Charles University grants SVV 260448 and GAUK 404316;
European Research Council, 7th Framework PIEF-GA-2013-622527, 
Horizon 2020 Marie Sklodowska-Curie grant agreement No. 700525 `NIOBE,' 
and
Horizon 2020 Marie Sklodowska-Curie RISE project JENNIFER2 grant agreement No. 822070 (European grants);
L'Institut National de Physique Nucl\'{e}aire et de Physique des Particules (IN2P3) du CNRS (France);
BMBF, DFG, HGF, MPG and AvH Foundation (Germany);
Department of Atomic Energy and Department of Science and Technology (India);
Israel Science Foundation grant No. 2476/17
and
United States-Israel Binational Science Foundation grant No. 2016113;
Istituto Nazionale di Fisica Nucleare and the research grants BELLE2;
Japan Society for the Promotion of Science,  Grant-in-Aid for Scientific Research grant Nos.
16H03968, 
16H03993, 
16H06492,
16K05323, 
17H01133, 
17H05405, 
18K03621, 
18H03710, 
18H05226,
19H00682, 
26220706,
and
26400255,
the National Institute of Informatics, and Science Information NETwork 5 (SINET5), 
and
the Ministry of Education, Culture, Sports, Science, and Technology (MEXT) of Japan;  
National Research Foundation (NRF) of Korea Grant Nos.
2016R1\-D1A1B\-01010135,
2016R1\-D1A1B\-02012900,
2018R1\-A2B\-3003643,
2018R1\-A6A1A\-06024970,
2018R1\-D1A1B\-07047294,
2019K1\-A3A7A\-09033840,
and
2019R1\-I1A3A\-01058933,
Radiation Science Research Institute,
Foreign Large-size Research Facility Application Supporting project,
the Global Science Experimental Data Hub Center of the Korea Institute of Science and Technology Information
and
KREONET/GLORIAD;
Universiti Malaya RU grant, Akademi Sains Malaysia and Ministry of Education Malaysia;
Frontiers of Science Program contracts
FOINS-296,
CB-221329,
CB-236394,
CB-254409,
and
CB-180023, and the Thematic Networks program (Mexico);
the Polish Ministry of Science and Higher Education and the National Science Center;
the Ministry of Science and Higher Education of the Russian Federation,
Agreement 14.W03.31.0026;
University of Tabuk research grants
S-1440-0321, S-0256-1438, and S-0280-1439 (Saudi Arabia);
Slovenian Research Agency and research grant Nos.
J1-9124
and
P1-0135; 
Agencia Estatal de Investigacion, Spain grant Nos.
FPA2014-55613-P
and
FPA2017-84445-P,
and
CIDEGENT/2018/020 of Generalitat Valenciana;
Ministry of Science and Technology and research grant Nos.
MOST106-2112-M-002-005-MY3
and
MOST107-2119-M-002-035-MY3, 
and the Ministry of Education (Taiwan);
Thailand Center of Excellence in Physics;
TUBITAK ULAKBIM (Turkey);
Ministry of Education and Science of Ukraine;
the US National Science Foundation and research grant Nos.
PHY-1807007 
and
PHY-1913789, 
and the US Department of Energy and research grant Nos.
DE-AC06-76RLO1830, 
DE-SC0007983, 
DE-SC0009824, 
DE-SC0009973, 
DE-SC0010073, 
DE-SC0010118, 
DE-SC0010504, 
DE-SC0011784, 
DE-SC0012704; 
and
the National Foundation for Science and Technology Development (NAFOSTED) 
of Vietnam under contract No 103.99-2018.45.

\bibliography{belle2}
\bibliographystyle{belle2-note}
\end{document}